\documentclass{mn2e}
\usepackage{graphicx}

\newif\ifAMStwofonts

\title[The peculiar variable V838 Mon]
  {The peculiar variable V838 Mon}
\author[S. Kimeswenger et al.]
  {S. Kimeswenger, C. Lederle, S. Schmeja, and B. Armsdorfer\\
  Institut f{\"u}r Astrophysik der Leopold--Franzens--Universit{\"a}t Innsbruck,
  Technikerstr. 25, A-6020 Innsbruck, Austria
}

\date{Accepted 2002 August 28.
      Received 2002 June 26; in original form 2002 May 2}

\pagerange{\pageref{firstpage}--\pageref{lastpage}} \pubyear{2002}

\begin{document}

\maketitle

\label{firstpage}

\begin{abstract}
V838 Mon underwent, after a first nova--like outburst in January
and a usual \hyphenation{decline}decline, a second outburst after
one month, and a third weak one again a month later. Moreover a
very small increase of the  temperature at the beginning of April
gives us a hint on  a physical process with a period of one month.
We obtained a $BVRI_{\rm C}$ time sequence and modelled the
photometric behaviour of the object. This leads us to the
conclusion that the interstellar foreground extinction has to be
\mbox{0\fm6 $\le$ E$_{B-V}$ $\le$ 0\fm8} and that the
quasi-photosphere had persistently unusually low temperatures for
nova--like systems. The photometry was used to follow the dramatic
changes of the expansion. While the appearing 10~$\mu$m excess can
be well described by the heating of material ejected during this
event, the IRAS emission near the location of the progenitor,
originates most likely from dust, which were formed during the
previous evolution of the object. Assuming that  the light echoes
are coming from circumstellar material, the distance is 640 to
680\ pc -- smaller than the 790\ pc given in Munari et al. (2002).
In our opinion V838~Mon and V4332~Sgr are manifestations of a new
class of eruptive variables. We do not count M31~RV to this class.

\end{abstract}


\section{Basic data}
V838 Mon was discovered by Brown et al. \shortcite{iauc7785_1} on
2002
 January 6. Pre--discovery images
show the target first on 2002 January 1. On February 2 the second
'outburst' \mbox{($\Delta_{\rm V}$ = 3\fm6)} was monitored in
detail (Kimeswenger et al. 2002a). Such a behaviour of an
extremely long pre-max halt or even a decline followed by a strong
outburst was, according to our knowledge, never observed before in
nova--like outburst, although HR Del and V723 Cas showed some
similarities (Terzan et al. 1974; D\"urbeck 1981, Munari et al.
1996).

Our data were obtained with the Innsbruck 60cm telescope
(Kimeswenger 2001; Kimeswenger \& Lederle 2002) and a direct
imaging CCD device. Until March 5 a CompuScope Kodak 0400 CCD
(4\farcm6 $\times$ 3\farcm1 field of view) was used. Thereafter
and in the night of February 8 an AP7p SITe 502e (8\farcm36
$\times$ 8\farcm36) was attached to the system.
\begin{figure}
\centering
\includegraphics[width=\columnwidth]{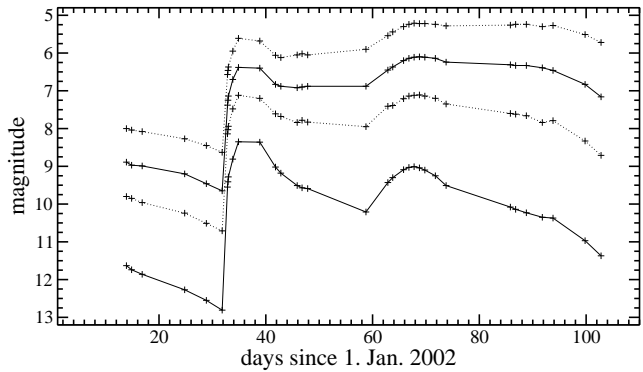}
\caption{The lightcurves (from top to bottom $I_{\rm C}$, $R$, $V$
and $B$) of V838 Mon between 2002 January 14 and 2002 April 13.
The errors (rms) are typically below 0\fm025. The lines do not
correspond to the real behaviour (especially around the first
maximum) but are given to connect the data of different bands
properly.} \label{lightcurve}
\end{figure}
\noindent 1018 images were taken in 31 nights between 2002 January
14 and April 13 using $BVR$ and $I_{\rm C}$ filters. Average
values for each day are given in Table~\ref{phot_tab} and the
resulting light curves shown in Fig.~\ref{lightcurve}. The
complete data table containing all individual measurements will be
published elsewhere (Kimeswenger et al. 2002b).

\begin{table}
\caption{Daily averaged photometry of V838 Mon} \label{phot_tab}
{\small
\begin{tabular}{lcccccc}
JD & day of  & V & B-V & V-R & R-I$_{\rm C}$ & V-I$_{\rm C}$ \\
   & 2002  &  & &  & &  \\
\hline
2452289.39 & 13.89 & 9.80 & 1.83 & 0.92 & 0.89 & 1.80 \\
2452290.36 & 14.86 & 9.85 & 1.89 & 0.88 & 0.93 & 1.81 \\
2452292.34 & 16.84 & 9.96 & 1.90 & 0.97 & 0.91 & 1.88 \\
2452300.30 & 24.80 & 10.24 & 2.03 & 1.04 & 0.93 & 1.97 \\
2452304.36 & 28.86 & 10.51 & 2.05 & 1.05 & 1.02 & 2.06 \\
2452307.33 & 31.83 & 10.71 & 2.10 & 1.06 & 1.02 & 2.08 \\
2452308.33 & 32.83 & 8.14 & 1.41 & 0.76 & 0.81 & 1.57 \\
2452308.42 & 32.92 & 8.03 & 1.38 & 0.78 & 0.79 & 1.57 \\
2452308.50 & 33.00 & 7.94 & 1.34 & 0.80 & 0.77 & 1.57 \\
2452309.32 & 33.82 & 7.48 & 1.33 & 0.78 & 0.75 & 1.53 \\
2452310.38 & 34.88 & 7.12 & 1.23 & 0.74 & 0.77 & 1.51 \\
2452314.37 & 38.87 & 7.20 & 1.16 & 0.80 & 0.72 & 1.52 \\
2452317.32 & 41.82 & 7.61 & 1.41 & 0.79 & 0.77 & 1.55 \\
2452318.32 & 42.82 & 7.68 & 1.50 & 0.80 & 0.77 & 1.56 \\
2452321.38 & 45.88 & 7.84 & 1.67 & 0.92 & 0.87 & 1.79 \\
2452322.29 & 46.79 & 7.78 & 1.79 & 0.89 & 0.88 & 1.76 \\
2452323.34 & 47.84 & 7.83 & 1.76 & 0.95 & 0.83 & 1.78 \\
2452334.26 & 58.76 & 7.95 & 2.26 & 1.07 & 0.98 & 2.05 \\
2452338.36 & 62.86 & 7.41 & 2.02 & 0.97 & 0.90 & 1.87 \\
2452339.27 & 63.77 & 7.39 & 1.91 & 1.03 & 0.93 & 1.95 \\
2452341.31 & 65.81 & 7.21 & 1.88 & 1.01 & 0.90 & 1.91 \\
2452342.31 & 66.81 & 7.14 & 1.89 & 1.00 & 0.90 & 1.90 \\
2452343.28 & 67.78 & 7.12 & 1.90 & 1.01 & 0.90 & 1.90 \\
2452344.28 & 68.78 & 7.11 & 1.93 & 1.01 & 0.88 & 1.90 \\
2452345.29 & 69.79 & 7.14 & 1.97 & 1.02 & 0.89 & 1.92 \\
2452347.26 & 71.76 & 7.20 & 2.05 & 1.06 & 0.90 & 1.96 \\
2452349.27 & 73.77 & 7.35 & 2.15 & 1.11 & 0.96 & 2.07 \\
2452361.31 & 85.81 & 7.60 & 2.48 & 1.29 & 1.05 & 2.34 \\
2452362.29 & 86.79 & 7.62 & 2.51 & 1.30 & 1.08 & 2.38 \\
2452364.30 & 88.80 & 7.66 & 2.57 & 1.33 & 1.09 & 2.42 \\
2452367.30 & 91.80 & 7.84 & 2.51 & 1.45 & 1.10 & 2.55 \\
2452369.30 & 93.80 & 7.79 & 2.58 & 1.33 & 1.19 & 2.52 \\
2452375.30 & 99.80 & 8.33 & 2.64 & 1.50 & 1.32 & 2.82 \\
2452378.30 & 102.80 & 8.71 & 2.66 & 1.56 & 1.44 & 3.00 \\
\hline
\end{tabular}}
\end{table}

\begin{figure}
\centering
\includegraphics[width=60mm]{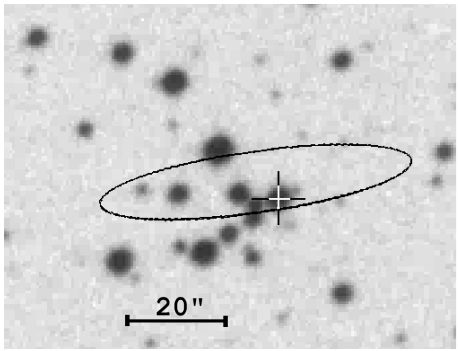}
\caption{The R-band SuperCOSMOS sky survey plate scan. The CCD
position is marked by the cross. The error of the position is
smaller than the thickness of the lines. The ellipse marks the
position of IRAS 07015-0346 (see Sec.~\ref{sec_ir}).}
\label{dssimage}
\end{figure}

Wagner et al. \shortcite{iauc7785_2} mention that the object on
the sky survey plate is a blend of at least two objects. To
identify unambiguously the progenitor, CCD frames from January and
SuperCOSMOS sky survey plate scans  were astrometrically
calibrated. The difference in position is below 30\ mas (see Fig.
\ref{dssimage}).

\begin{figure}
\centering
\includegraphics[width=\columnwidth]{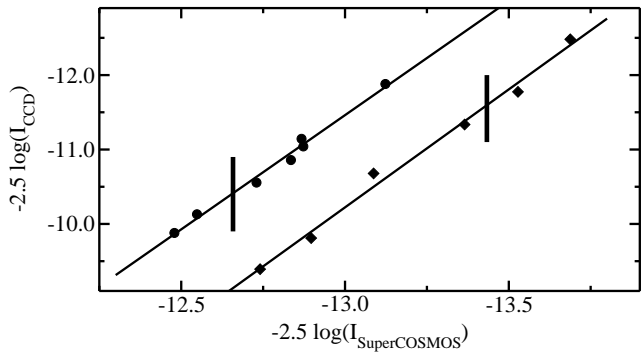}
\caption{The calibration of the SuperCOSMOS sky survey plate scans
(circles: $B$-band; diamonds: $R$-band) using deep CCD frames. The
bars mark the positions of the progenitor for each band.}
\label{dss}
\end{figure}

Furthermore a set of deep $B$ and $R$ images were taken to
calibrate the stars surrounding V838 Mon on the SuperCOSMOS frames
(Fig.~\ref{dss}). Using the colour corrections from H{\"o}rtnagl
et al. (1992), we find the pre--outburst magnitudes
$R=14\fm56\,\pm\,$0\fm10 and $B=15\fm87\,\pm\,$0\fm10. However,
the ($B-R$) may be used only with care, since the plate epochs
differ by 6.13 years. If the object was already variable, the
colour is not meaningful. Our values differ from those given by
Munari et al. (2002). This difference in $R$ is higher than
expected from the different filter sets used ($R$ vs. $R_{\rm
C}$). As they give no information on the type of plate scan used
in their work (DSS, SuperCOSMOS, \dots) we are unable to compare
the results. No information also is given there, how they
transformed POSS E and ESO R ($\lambda_{\rm eff} = 645$nm).
Moreover their conclusion that a red object \mbox{[(B-V) = 0\fm9}
in their calibration] has the same magnitude on POSS I
($\lambda_{\rm eff} = 410$nm) and SERC $J$ ($\lambda_{\rm eff} =
480$nm) is confusing. They also include the $B$-band to the fit of
the spectral energy distribution (SED), although this band is
known to be strongly affected by non grey opacities at
temperatures below \mbox{10\,000~K}. This leads to a lower
temperature and thus later on to a lower interstellar extinction.

\section{interstellar extinction, temperatures and radii}
\begin{figure}
\centering
\includegraphics[width=\columnwidth]{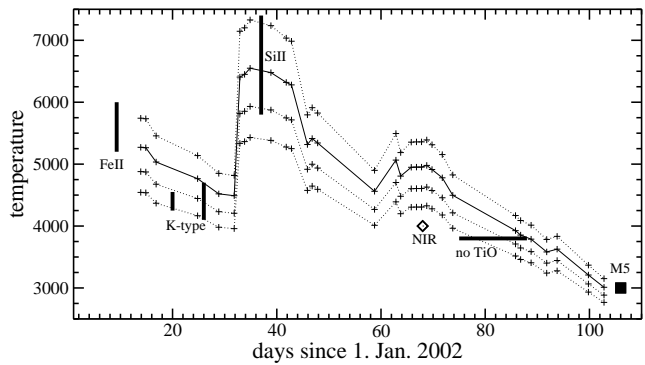}
\caption{The evolution of the photosphere temperature as function
of the assumed foreground extinction (from top to bottom E$_{\rm
B-V}$ = 0\fm8, 0\fm7, 0\fm6 and 0\fm5). The thick marks represent
spectroscopic boundary conditions from literature (see text).}
\label{TLbol}
\end{figure}

As the early spectra did not show strong spectral features
affecting  our wide band photometry (Wagner et al. 2002, Della
Valle \& Iijima 2002, Geballe et al. 2002a, Zwitter \& Munari
2002, Lynch et al. 2002) a blackbody approximation is applicable
at $\lambda > 0.5\,\,\mu$m. We used the flux $F$ for a given epoch
$i$ at a wavelength $\lambda$
\begin{equation}
F_{\lambda,i} \propto {R_i^2 \over \lambda^5} \,\, {1 \over
{\exp\left({c_2 \over {\lambda T_i}}\right) - 1} } \,\,
10^{-0.4\,{\rm A}_\lambda}
\end{equation}
where $R_i$ is the radius of the photosphere, $c_2 = 0.0144
\,\,\mbox{\rm m K}$, $T_i$ the Planck colour temperature and
A$_\lambda$ the interstellar foreground extinction. We applied
this to the V and the I$_{\rm C}$ bands, as they are less affected
by lines than B and R, to define a colour--dependent quantity
\begin{equation}
{F_{V,i} \over F_{I,i}} = C_{i}^{VI} = \left({\lambda_I \over
\lambda_V}\right)^5 \,\,{ {\exp\left({c_2 \over {\lambda_I
T_i}}\right)-1} \over {\exp\left({c_2 \over {\lambda_V
T_i}}\right)-1}} \,\, 10^{-0.4992 \,{\rm E}_{\rm B-V}}
\end{equation}
using the interstellar extinction law (Mathis et al. 1977, Koller
\& Kimeswenger 2001). This leads us to a complete sequence for the
temperature as a function of the interstellar extinction
(Fig.~\ref{TLbol}). The relative radii $R_i/R_0$, obtained by the
combination of the equations, are independent from the extinction.
These radii show a smooth evolution with time that was fitted by
means of an empirical function. Using the radius $R_0$ from
January 14 as reference we thus can approximate the radius by
\begin{equation}
R(d) = 0.6212 \times R_0 \,\, \exp(0.0273 \times d),
\end{equation}
with $d$ being the day--number of the year 2002. Spectroscopic
observations give us now boundaries for the extinction. While the
spectroscopy of Wagner et al. (2002) around $d = 10$ (namely FeII
absorption lines but no SiII) suggests \mbox{5100 $< T <$ 6000\ K}
and thus \mbox{0\fm7 $\le$ E$_{\rm B-V}$ $\le$ 0\fm9}, the
findings of Zwitter \& Munari \shortcite{iauc7812_3}, $d = 26$ and
the spectra of Fujii (2002) from $d = 19$ yield to \mbox{0\fm5
$\le$ E$_{\rm B-V}$ $\le$ 0\fm7}. The occurrence of CaII lines
before and SiII lines just after the February 2 outburst (Morrison
et al. 2002), confines via the Saha equation (assuming that the
density did not change significantly within those few days) the
increase of temperatures to at least 1:1.36. As the P-Cygni
profiles around that time are similar, we have to assume, that the
radiation originates from the same region of the shell. So we get
a lower limit of \mbox{0\fm60 $<$ E$_{\rm B-V}$}. The SiII
disappeared just after a few days. Thus the value should be rather
near to that boundary. This gives an estimated density of
\mbox{$N_e \approx 10^{19}$\ m$^{-3}$}. The temperature from the
NIR continuum (Geballe et al. 2002b), although not very accurate
at the beginning Raleigh-Jeans domain of the blackbody, leads to a
lower extinction. The non-appearance of TiO bands before April
limits the temperature in March and again gives \mbox{0\fm7 $\leq$
E$_{\rm B-V}$} (Gavin 2002, Morata \& Morata 2002). Furthermore,
when referring to an extinction of  \mbox{0\fm7 $\le$ E$_{\rm
B-V}$ $\le$ 0\fm8}, our results for April fit also very well to
the M5 spectral temperature given by Rauch \& Kerber
\shortcite{iauc7886_2}.
 Thus we assume an interstellar
extinction of \mbox{E$_{B-V}$ $\approx$ 0\fm7} for the following
calculations.

The resulting temperatures were used to fit a SED also for the
dates when NIR data  by Bailyn \shortcite{iauc7791_3}, Geballe et
al. (2002a, 2002b), Lynch et al. \shortcite{iauc7829_1} and Hinkle
et al. \shortcite{iauc7834_1} were available. The results for all
bands in the V to K range deviate less than 15 percent. Altogether
these facts suggest, a simple blackbody model in fact may be
applied here.

\section{The Distance}

Using expansion and diameter of the light echo reported by Henden
et al. \shortcite{iauc7859_1} and the size from the HST
observations (Bond et al. 2002a), we obtain, assuming spherical
geometry, a distance of 640 to 680\ pc. Munari et al. (2002)
obtained a higher value of 790\ pc from a different analysis of
only the Henden et al. data. The larger distance scale is
supported by the measurements of Orio et al. (2002). But the
latter are using red bands and are thus less sensitive at the
outer edge. Thus the HST calibration, and thereafter the shorter
distance, seems more reliable to us. Using the WHT image (Figure 2
in Munari et al. 2002) we also obtain 650\ pc and not 790\ pc as
stated there.

Recently Bond et al. (2002b) reported new HST observations from
2002 May 20. They conclude from the radius of 20\arcsec a minimum
distance of 2.5~kpc, using the weak peak on March 11 as outburst
date. But the light echoes are reported first already mid of
February. We thus assume that the February 2 outburst started the
illumination of circumstellar material. With respect to the
possibility of a slab of material in front of the star, as used in
the estimate by Bond et al. (2002b), we like to refer to the
detailed discussion in Munari et al. (2002). There it is already
shown that such a geometry is very unlikely. Taking into account
our assumptions, the USNO and WHT images of Munari et al., the
first HST images and the marginal growth of the structure from
April 30 until May 20, we conclude that the light has reached the
outer boundary of the dust shell. Thus the shell has a size of
0.1~pc. The inner edge has grown since end of April. This is due
to the rapid decrease of brightness just 2 weeks before (16 light
days are 5\arcsec at a distance of 650 pc).

Although the distance is somewhat sensitive to the geometry of the
dust around the object, we adopt 650\ pc here for the further
discussion yielding to \mbox{M$_V \approx -4\fm5$} at maxi\-mum
visual light.

\section{IRAS 07015-0346 and the 10 micron excess of V838 Mon}
\label{sec_ir}

\begin{figure}
\centering
\includegraphics[width=\columnwidth]{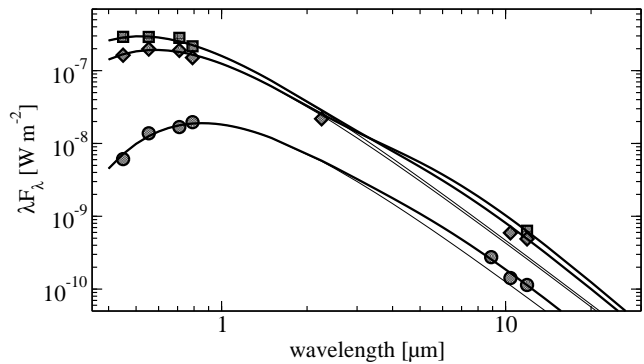}
\caption{ The MIR excess modelled with a constant condensation
temperature (see text). Thick lines: the total model photosphere +
dust cloud; thin lines: pure photosphere; symbols: circles January
29, boxes February 8, diamonds February 12. } \label{MIR}
\end{figure}

The model of the visual to near infrared photometry gives us a 10\
$\mu$m flux excess (Fig.~\ref{MIR}). The 10\ $\mu$m excess was
already obvious before the second outburst and increased steadily
(K{\"a}ufl et al. 2002). Using the grey dust temperature (Leene \&
Pottasch 1988, Evans 1994) and assuming an AGB-- or C--star like
condensation model with an average \mbox{$T_{\rm C} = 750$\ K}
(Dominik et al. 1993) the condensation is possible at $r = 3
\times 10^{11}$\ m in January and $r = 1.4 \times 10^{12}$\ m in
February. Assuming a constant dust formation rate in January and
again a constant but higher one in February, we find $3.5 \times
10^{-10}$\ M$_\odot$/yr and $2 \times 10^{-6}$\ M$_\odot$/yr,
respectively, leading to reasonable values of 150 to 350\
km\,s$^{-1}$ for the mass loss wind. The material, causing the
light echoes starting in a distance of about \mbox{$3 \times
10^{14}$\ m} and reaching out to \mbox{$2 \times 10^{15}$\ m}
(measured by us on the HST images) cannot be the origin of the MIR
excess.

Kato et al. \shortcite{iauc7786_3} stated that the position of the
target and that of the IRAS PSC source 07015-0346 correspond very
well. The position is far inside the light echoes. Using the space
density of IRAS sources within 100 square degrees, a chance for a
pure coincident superposition is about 1:50\,000. As there are no
large CO clouds known in that region, the IRAS PSC source is not
very likely connected to a cold background molecular cloud. The
source is best fitted with a 35$\pm$3 K blackbody spectrum. At a
distance of 650\ pc we obtain a dust mass of a few times
\mbox{10$^{-4}$ M$_\odot$}. A single previous event like the one
this year cannot form such a cloud but a continuous phase of
massive mass loss by a stellar wind may produce such an
environment. We assume that the IRAS PSC source is the dust
causing the light echoes. Our dust code NILFISC (Koller \&
Kimeswenger 2001) and the mass estimate given above, leads to a
contribution of the circumstellar material to the total extinction
of E$_{\rm B-V}^{circum.} \le 0\fm1$. Assuming a typical albedo of
0.6, this leads to a total integrated reflected light of
approximately 15\ percent of the unreddened star. This is, within
a factor of 3, in fact the brightness of the halo on the
(polarimetric) HST images April 30, if integrating the nearly
constant visual magnitude beginning with the February outburst.

\section{Discussion}

\paragraph*{CV or Nova--like outburst ?\medskip\newline}

Using, although this may be not applicable for this peculiar
object, the steady decline after the first primary maximum 2002
January 10, the thick wind models of Hachisu \& Kato
\shortcite{HaKa} suggest a low mass WD with about 1 M$_\odot$. The
optical and $JHK$ (2MASS) pre-outburst photometry, even if the use
of the colours might be dangerous, is either fitted by a single
7800$\pm$300\ K blackbody or better by a composite of a hot
accretion disk with a \mbox{4500\ K} companion. At a distance of
650\ pc the absolute visual magnitude would be around +3\fm6. A
single early F type main sequence star would fit to these values.
Using the better fitting double component model and using the
accretion disk parameters from Webbink et al. \shortcite{We87} we
obtain a K2-4 main sequence star and a very reasonable accretion
rate of $2 \times 10^{-8}$~M$_\odot$~yr$^{-1}$. Due to the higher
extinction and the different values of the photometry for the
progenitor, we do not need an unusual subluminous star as it is
used by Munari et al. (2002). The temperatures during outburst are
too low for such a classical scenario. Although the models of M31\
RV (Iben \& Tutukov 1992) allow such a red cold outburst in small
period CV with a cold low mass WD, it leads to a higher outburst
luminosity (M$_{\rm V} \approx -10$).  Munari et al. (2002)
discuss this scenario with respect to a  thermonuclear runaway of
the complete WD. However M31 RV underwent at least two recurrent
outbursts within the rather small timescale of 20 years (Sharov
1990). We thus infer  that the nova/CV scenario have to be
excluded

\paragraph*{A late helium flash post-AGB ?\medskip\newline}

This scenario (V605 Aql, V4334\ Sgr, \dots) was discussed several
times in different IAU circulars.  Both, V605 Aql as well as V4334
Sgr, showed a 6000-7000\ K photosphere throughout more than one
year. The light curves were smooth. For V4334~Sgr (outburst in
February 1996) we know now, that it stayed at constant bolometric
luminosity for at least seven years. V605~Aql has the same
bolometric luminosity even 85 years after its outburst. Even if we
adopt a much larger distance and thus a more luminous scenario, we
should have a much slower evolution after the outburst. Also the
lack of strong carbon overabundance in the spectra clearly rules
out this model.

\paragraph*{A twin of V4332\ Sgr ?\medskip\newline}

V4332\ Sgr underwent a similar outburst in 1994 (Martini et al.
1999 and references therein). Its nature is unclear up to now.
While the evolution of the temperature and the expansion of an
opaque massive shell and the luminosity at maximum look much the
same, V4332\ Sgr did not show multiple visual maxima like V838\
Mon. This statement, as well as the statement that the ascent to
maximum lasted 200 days (Martini et al. 1999), have to be used
with care. V4332~Sgr was not observable before the discovery
(daytime object). If we compare the light curve and the
temperature sequence of V838~Mon with those of V4332~Sgr, we see a
good agreement, starting mid of March 2002. It is thus possible,
that the earlier features did also exist there. They were simply
not observed. A single thermonuclear event, as discussed for
V4332\ Sgr, still might cause such a behaviour. Also the density
of \mbox{$N_e \approx 10^{14}$ m$^{-2}$} (Martini et al. 1999) for
the late stages fits well to our finding above, if we take into
account the 15 times larger radius, compared to the date when we
estimated the density. But this does not agree with the main
sequence progenitor. Why should a main sequence star undergo a
shell flash? Low mass blue horizontal branch (BHB) stars evolve
directly to a WD without passing through a luminous AGB phase
(Heber et al. 1997). They are stated to be the progenitors of low
mass WDs ($\approx 0.3 M_\odot$) without planetary nebula. Those
stars, at the end of the He-core burning in transition to He-shell
burning, pass the main sequence at a similar position in the
HR-diagram. This would be a new possible scenario similar to a
late He-flash for normal mass progenitors passing through the AGB.

\bigskip

Although it is too early to draw final conclusions, the
photometric behaviour, the temperature and the bolometric
luminosity most likely make V838~Mon a twin of V4332~Sgr. Those
two objects seem to form a new class of eruptive variables. Unlike
Munari et al. (2002), we do not include M31~RV in this group. The
photometric and spectroscopic behaviour and especially the
luminosity of this object seem to differ significantly. We have to
think about possible physical processes causing a star near the
main sequence to ignite eruptive thermonuclear events.\\
Further observations at all wavelengths of the up to now unstudied
remnant of V4332~Sgr and the new remnant expected after V838~Mon
emerges from its solar conjunction are urged.

\section*{Acknowledgments}
We thank the head of the institute, R. Weinberger, for the
unlimited access to the facilities of the new university
observatory. We also thank the referee for his helpful remarks
during revision of the paper. This research used data originating
from the SuperCOSMOS digitization of UK Schmidt plate material.

\label{lastpage}
\end{document}